\documentclass{article}
\usepackage{spconf}
\usepackage{amsmath,amssymb,amsfonts}
\usepackage{graphicx}
\usepackage{algorithm}
\usepackage{algpseudocode}
\usepackage[table,dvipsnames,x11names]{xcolor}
\usepackage{stfloats}
\usepackage[
    colorlinks=true,         
    linkcolor=MidnightBlue, 
    citecolor=OliveGreen,    
    filecolor=magenta,       
    urlcolor=BlueViolet      
]{hyperref}
\usepackage[skip=5pt]{caption}
\usepackage{enumitem}
\setlist{nosep, leftmargin=14pt}
\usepackage{hyperref}
\usepackage[capitalize, nameinlink]{cleveref}
\usepackage{bm}
\usepackage{tabularray}
\UseTblrLibrary{booktabs}
\usepackage{multirow}   
\SetTblrStyle{caption}{hang=0pt,halign=j}

\usepackage[backend=bibtex,
sorting=none,
doi=false,
url=false,
isbn=false,
style=ieee,
]{biblatex}
\usepackage[range-units=single]{siunitx}
\usepackage{lipsum}
\usepackage{soul}
\usepackage{tikz}
\usetikzlibrary{calc}

\usepackage{amsthm}
\usepackage{mathtools}
\usepackage{xcolor}

\newtheorem{prop}{Proposition}

\usepackage{xfrac}
\usepackage{bm}

\title{Robust Plug-and-Play Methods for Highly Accelerated Non-Cartesian MRI Reconstruction}
\name{Pierre-Antoine Comby\(^{\star \dagger}\) \qquad Benjamin Lapostolle\(^{\star \dagger}\)  \qquad Matthieu Terris\(^{\dagger}\) \qquad Philippe Ciuciu\(^{\star \dagger}\)}

\address{
\(^{\star}\)CEA, Joliot, NeuroSpin, Universit\'e Paris-Saclay, F-91191 Gif-sur-Yvette, France \\
\(^{\dagger}\)Inria, MIND, Universit\'e Paris-Saclay, F-91120 Palaiseau, France\\
}
\begin{document}
\ninept
\maketitle

\begin{abstract}
Achieving high-quality Magnetic Resonance Imaging (MRI) reconstruction at accelerated acquisition rates remains challenging due to the inherent ill-posed nature of the inverse problem. Traditional Compressed Sensing (CS) methods, while robust across varying acquisition settings, struggle to maintain good reconstruction quality at high acceleration factors~(\(\geq 8\)). Recent advances in deep learning have improved reconstruction quality, but purely data-driven methods are prone to overfitting and hallucination effects, notably when the acquisition setting is varying.
Plug-and-Play~(PnP) approaches have been proposed to mitigate the pitfalls of both frameworks. In a nutshell, PnP algorithms amount to replacing suboptimal handcrafted CS priors with powerful denoising deep neural network~(DNNs).
However, in MRI reconstruction, existing PnP methods often yield suboptimal results due to instabilities in the proximal gradient descent (PGD) schemes and the lack of curated, noiseless datasets for training robust denoisers.
In this work, we propose a fully unsupervised preprocessing pipeline to generate clean, noiseless complex MRI signals from multicoil data, enabling training of a high-performance denoising DNN. Furthermore, we introduce an annealed Half-Quadratic Splitting~(HQS) algorithm to address the instability issues, leading to significant improvements over existing PnP algorithms. When combined with preconditioning techniques, our approach achieves state-of-the-art results, providing a robust and efficient solution for high-quality MRI reconstruction.
\end{abstract}

\section{Introduction}

The quest for increasing resolution and shorter acquisition times in magnetic resonance imaging (MRI) has led to the development of more advanced acquisition strategies.
In this context, accelerated multicoil acquisition has become a standard practice in clinical settings, allowing faster scans while maintaining image quality. However, as acceleration factors increase, less k-space samples are collected and image reconstruction becomes increasingly challenging due to the ill-posed nature of the underlying inverse problem. 

To address this issue, reconstruction techniques based on Compressed Sensing (CS) theory have been widely adopted~\cite{lustig2007sparse}. 
Despite their success and the rise of a new generation of variable density undersampling patterns~\cite{lazarus2019sparkling,chaithya2022optimizing}, CS-based methods often deliver suboptimal image quality, particularly at higher acceleration factors, where the ill-posedness of the problem is most severe.
Recently, driven by the greater availability of public MRI data sets and computational power, deep learning methods have propelled image reconstruction quality to new heights~\cite{zbontar2018fastmri,hammernik2018learning,muckley2021results,ramzi2022nc, pramanik2023memory}.
However, their performance is highly dependent on the acquisition setup (\emph{e.g.} image resolution, sampling trajectory, acceleration factor, SNR of the acquisition, etc.) and are prone to overfitting and hallucinations.

Therefore, hybrid methods have emerged, which combine the capabilities of both the deep learning approach and the robustness of traditional CS methods.
In particular, the Plug-and-Play (PnP) framework~\cite{venkatakrishnan2013plug, huraultgradient,pesquet2021learning} has been shown to be a powerful approach to image restoration tasks and has been successfully applied to MRI reconstruction in various settings~\cite{wei_tfpnp_2021,ahmad_plug_2020, hong_provable_2024, fatania_plug-and-play_2022}. However, the performance of PnP algorithms is strongly dependent on the quality of the learned prior, and training with appropriate data is key.
This is problematic in applications dealing with indirectly acquired signals, such as in MRI, where a reconstruction algorithm is necessary to generate the reference data, potentially introducing biases, artifacts, or inaccuracies into the training data. 
Furthermore, the convergence properties of the PnP algorithm can be challenging to ensure~\cite{hong_provable_2024, terris2024equivariant}. 

In this work, we propose to bypass both these issues by (i) denoising the multicoil fastMRI dataset in a fully unsupervised way and, (ii) applying an annealed half-quadratic-splitting~(HQS) algorithm yielding fast reconstruction with state-of-the-art performance, while guaranteeing stable convergence.

\section{Plug-and-Play algorithms for MRI}
\subsection{MR image reconstruction}
In an ideal~(i.e.\ artifact-free) setting, MR imaging consists of collecting the Fourier transform of the spatial distribution of nuclear magnetization in the organ of interest. In a multicoil~(say \(L\)) acquisition setup, the signal measured by each coil \(\ell\) surrounding the organ can be modelled as:
for \(\ell \in \{1,\hdots, L\}\),
\begin{equation}
  \label{eq:MRI_multicoil}
  y_{\ell} = \mathcal{F}_\Omega\mathcal{S}_{\ell}x + e_{\ell},
\end{equation}
where \(x\in \mathbb{C}^{N}\) is the image to reconstruct, \(\mathcal{F}_\Omega\) is the (potentially non-uniform) Fourier transform operator depending on a sampling trajectory \(\Omega\), \((\mathcal{S}_{\ell})_{1\leq\ell\leq L}\) are the complex-valued spatial sensitivity maps (or S-maps), and \(e_{\ell}\) is the realization of some random noise. The S maps are smooth low-frequency images and can be estimated from the central frequencies of \(y_\ell\). The goal of the reconstruction is then to recover the image \(x\) from the data \((y_{\ell})_{1\leq \ell \leq L}\).

\subsection{Plug-and-play algorithms}

Traditional image reconstruction algorithms propose to solve the ill-posed problem~\eqref{eq:MRI_multicoil} by reformulating it as a minimization problem of the form
\begin{equation}
    \widehat{x} = \arg\min_{x\in\mathbb{C}^{N}} \frac{1}{2} \sum_{\ell=1}^{L} \| \mathcal{F}_\Omega\mathcal{S}_{\ell}x - y_{\ell}\|^2 + \lambda r(x),
\end{equation}
where \(r\) is a convex function and \(\lambda>0\) a regularization parameter. In this context, the choice of an appropriate regularizer \(r\) is paramount to ensure good reconstruction quality to counterbalance the ill-posedness of~\eqref{eq:MRI_multicoil}. Denoting \(f(x)= \frac{1}{2} \sum_{\ell=1}^{L} \| \mathcal{F}_\Omega\mathcal{S}_{\ell}x - y_{\ell}\|^2\), the solution \(\widehat{x}\) is then obtained by iterating a proximal algorithm~\cite{combettes2011proximal}, a classical instance being the proximal gradient algorithm
\begin{equation}
\label{eq:PGD}
x_{k+1} = \operatorname{prox}_{\gamma \lambda r}(x_k - \gamma \nabla f(x_k)),
\end{equation}
where we recall that \(\operatorname{prox}_{\gamma \lambda r}(x) \stackrel{\text{def}}{=} \underset{u}{\text{argmin}}\,\, \gamma \lambda r(u) + \frac{1}{2}\|x-u\|_2^2\).
Plug-and-Play~(PnP) algorithms~\cite{venkatakrishnan2013plug, pesquet2021learning, huraultgradient} propose to replace the proximal operator in~\eqref{eq:PGD} with a denoising neural network \(\operatorname{D}_{\sigma}\), where \(\sigma>0\) is the denoising power.
There, \(\operatorname{D}_\sigma\) accounts for an implicit prior more expressive than standard regularizers \(r\), and the noise level \(\sigma\) plays a similar role as \(\lambda\). PnP algorithms have been shown to strongly outperform their traditional counterparts in image restoration~\cite{zhang2021plug}, MRI~\cite{hong_provable_2024}, and astronomical imaging~\cite{dabbech2022first} to name a few.

PnP algorithms are owing their popularity to several advantages. Firstly, the sophisticated implicit prior \(\operatorname{D}_\sigma\) is trained independently from the acquisition strategy in\eqref{eq:MRI_multicoil}: If the under-sampling pattern \(\Omega\) changes, the same \(\operatorname{D}_\sigma\) can be used for the reconstruction, which constitutes a significant advantage over unrolled algorithms. Secondly, while the resulting algorithms may be unstable, annealing schemes~\cite{zhang2021plug} allow us to stabilize the algorithm while drastically improving the reconstruction.

As a consequence, PnP algorithms have been used for MRI reconstruction in a number of works~\cite{chand_multi-scale_2023, hong_provable_2024, wei_tfpnp_2021}.
Much of these works outline the importance of using convergent reconstruction algorithms. In this work, we are rather leveraging reconstruction algorithms with annealing strategies.

\subsection{Complex-valued denoisers for MRI}

At the algorithmic level, the proximal operator in~\eqref{eq:PGD} takes complex variables as input. Similarly, the denoiser used in the PnP algorithm will be faced with complex images with real and imaginary parts once a multicoil combination has been performed at each gradient step. Therefore, PnP algorithms for multicoil MRI data need to rely on denoisers that handle complex data.

Moreover, the quality of the reconstruction of the PnP algorithms is highly dependent on that of the denoiser at hand~\cite{terris2023plug}.
However, in the MRI context, since the image is indirectly acquired and often requires preprocessing, either linear in the parallel imaging setting or nonlinear, such as~\eqref{eq:PGD}, no data set of ground truths (noiseless) is available, and only noisy datasets~(e.g. fastMRI) are available.
Several works have proposed to denoise MR images~\cite{chung2022mr, pfaff2023self}, but to our knowledge, none applied it to PnP algorithms. In this work, we denoise the fastMRI dataset in an unsupervised way to train our denoisers.

\section{Proposed approach}

In this section, we detail our proposed approach. 
We first introduce our strategy for producing a clean dataset of complex-valued images from multicoil MR acquisitions, which will serve as a basis for training our denoising DNN. We next introduce the proposed preconditioned HQS algorithm and the choices considered for the preconditioning matrix.

\subsection{Denoised dataset generation}
\label{sect:dataset_gen}
We rely on the fastMRI dataset~\cite{zbontar2018fastmri} for our study, which provides only fully
sampled multicoil data. 
To reconstruct a single complex image from multicoil measurements, we propose a virtual coil combination method. This approach transforms each coil’s k-space data into the image domain using FFT, aligns the phase information, and combines the coil images into a single complex image \(x_c\)~\cite{parker2014phase}.
We apply this operation to the full multicoil training dataset of fastMRI
\(\{(y^1_{1}, \hdots, y^1_{L}), \hdots, (y^N_{1}, \hdots,y^N_{L})\}\),
yielding a new dataset of \(N\) complex images \(\{x_c^1, \hdots, x_c^N\}\).

The dataset still contains residual acquisition noise, which can be significant in some cases. To address this issue, we propose using these complex, noisy images to train a restoration neural network, \(f_\theta\), in an unsupervised manner with the \texttt{Neighbor2Neighbor} loss~\cite{huang2021neighbor2neighbor}. This approach allows us to preprocess the data without relying on clean, noiseless samples. Specifically, for a given noisy sample \(x_c\) the training loss is computed as follows:
\begin{align}\label{equ::psen2n_rerm}
 \mathcal{L}(x) &= \left\lVert f_\theta(g_1(x_c)) - g_2(x_c)\right\rVert_2^2 \\
& + \eta \left\lVert f_\theta(g_1(x_c)) - g_2(x_c) - (g_1(f_\theta(x_c)) - g_2(f_\theta(x_c))) \right\rVert_2^2, \nonumber
\end{align}
where \(g_1\) and \(g_2\) are sub-sampling operations and \(\eta>0\) a regularization parameter.

After training, we apply the learned model to the full dataset of combined images \(\{x_c^1, \hdots, x_c^N\}\), yielding a new training dataset of \(N\) denoised, complex images \(\mathcal{D}_\text{train} = \{x^1, \hdots, x^N\}\) that will serve as a basis for training our implicit prior \(\operatorname{D}_\sigma\).

\subsection{Unsupervised preprocessing of multicoil data}
\begin{figure}
    \centering
    \includegraphics[width=0.4\textwidth]{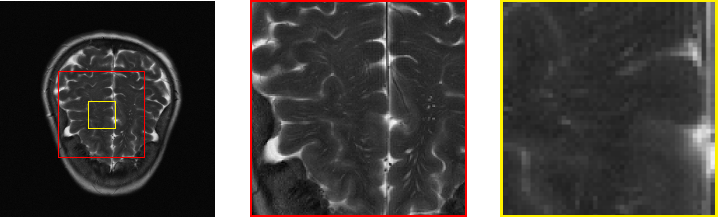} \\
    \includegraphics[width=0.4\textwidth]{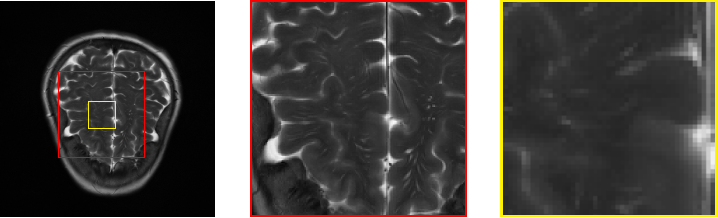}
    \caption{Effect of the unsupervised preprocessing on the training dataset. Top row shows a non-preprocessed sample of the fastMRI multicoil dataset, obtained with virtual coil combination procedure described in~\cite{parker2014phase}; bottom row shows the preprocessed image.}
    \label{fig:preprocessing_illust}
\end{figure}

We apply the procedure described in Section~\ref{sect:dataset_gen} to the fastMRI brain multicoil dataset. We choose a DRUNet model~\cite{zhang2021plug} for \(f_\theta\) in~\eqref{equ::psen2n_rerm}, with an input and output convolutions channel number set to 2 for real and imaginary parts, without conditioning on the noise level. Training is performed with a batch of size 10 and on randomly extracted patches of size 128\(\times\)128. We use the Adam optimizer with learning rate \(10^{-4}\), and set \(\eta=2.0\) in~\eqref{equ::psen2n_rerm}. We then preprocess the full training dataset with \(f_\theta\). An illustration of the resulting preprocessed dataset is given in Fig.~\ref{fig:preprocessing_illust}.
Notice that the resulting image shows much lower residual noise, while maintaining a high resolution.

\subsection{Proposed algorithm}
We propose to investigate several variants of PnP algorithms.
We first study the standard PnP version of the PGD algorithm~\eqref{eq:PGD} where the proximal operator is replaced by a denoising neural network \(\operatorname{D}\), writing
\begin{equation}
\label{eq:pnp_pgd}
\tag{PnP-PGD}
\begin{array}{l}
u_k = x_k - \gamma P \nabla f(x_k) \\
x_{k+1} = \operatorname{D}_\sigma(u_k),
\end{array} 
\end{equation}
where \(P\) is a preconditioning matrix, and \(\gamma>0\) and \(\sigma>0\) are stepsizes and noise levels, respectively.
While the use of \(P=\operatorname{Id}\) remains predominant in the image restoration literature, recent works have shown that using carefully tailored preconditioners strongly improves the reconstruction quality in MRI~\cite{hong_provable_2024}.

Following the DPIR approach~\cite{zhang2021plug}, we propose to extend the preconditioning of~\cite{hong_provable_2024} and use a half-quadratic splitting algorithm coupled with an annealing strategy summarized hereafter:
\begin{equation}
\label{eq:dpir}
\tag{PnP-HQS}
\begin{array}{l}
u_k = \operatorname{prox}_{\gamma_k f}^P(x_k) \\
x_{k+1} = \operatorname{D}_{\sigma_k}(u_k),
\end{array} 
\end{equation}
where \(\operatorname{prox}_{\gamma f}^P\) is the proximity operator of \(\gamma f\) in the metric induced by \(P\),
and \((\gamma_k)_{1 \leq k\leq K}\) (resp. \((\sigma_k)_{1\leq k\leq K}\)) are sequences of stepsizes (resp. noise levels). We underline that, in the case of \(P=\operatorname{Id}\),~\eqref{eq:dpir} reduces to the DPIR algorithm from~\cite{zhang2021plug}.
Importantly, we point out that despite their similarities,~\eqref{eq:pnp_pgd} and~\eqref{eq:dpir} differ fundamentally, as illustrated by the following result:
\begin{prop}
Assume that there exists a convex function \(g\) such that \(\operatorname{D}_\sigma= \operatorname{prox}_g\). Furthermore, assume that \(P = \operatorname{Id}\) and that for all \(k\), \(\gamma_k = \gamma\) and \(\sigma_k = \sigma\) in~\eqref{eq:dpir}. Then:
\begin{enumerate}
  \item~\eqref{eq:pnp_pgd} converges to \(x^* = \underset{x}{\text{argmin}}\,\, \gamma f(x) + g(x)\)~\cite{pesquet2021learning}.
  \item~\eqref{eq:dpir} converges to \(x^* = \underset{x}{\text{argmin}}\,\, \gamma f(x) + \prescript{1}{}g(x)\), where \(\prescript{1}{}g\) denotes the Moreau envelope~\cite{bauschke2017correction} of \(g\).
\end{enumerate}
\end{prop}

\noindent We note that the two algorithms converge to different points. Although the minimization problem associated with~\eqref{eq:pnp_pgd} seems more natural than that of~\eqref{eq:dpir}, empirical results show that the latter tends to outperform the former. Secondly, the assumption that \(\operatorname{D} = \operatorname{prox}_g\) for some convex function \(g\) is well-known to not hold in practice, resulting in non-convergent and potentially unstable algorithms.

Therefore, numerous works have proposed to enforce the convergence of the algorithm through Lipschitz constraints~\cite{pesquet2021learning, huraultgradient} or through preconditioning strategies~\cite{hong_provable_2024}.
Inspired by~\cite{zhang2021plug}, we propose to set \((\gamma_k)_{1 \leq k\leq K}\) and \((\alpha_k)_{1 \leq k\leq K}\) as decreasing sequences according to the rules \(\sigma_k = \sigma_0 \xi^\delta\) and \(\gamma_k = \lambda \sigma_k\), where \(\delta, \lambda > 0\) and \(0<\xi<1\) are tunable hyperparameters. This can be interpreted as a way to force convergence of \((x_k)_{1\leq k\leq K}\), thereby circumventing the instability issues witnessed with~\eqref{eq:pnp_pgd}.

\subsection{Preconditioning matrices}

In~\cite{hong_provable_2024}, the authors show that introducing preconditioning improves both the stability of the algorithm and the quality of the reconstruction.
We propose to extend their approach to the case of the HQS algorithm~\eqref{eq:dpir}.
In our case, the preconditioning is applied to the metrics of the proximal operator of the data-fidelity term \(\operatorname{prox}_{\gamma f}^P(x) \stackrel{\text{def}}{=} \underset{u}{\text{argmin}}\,\, \gamma f(u) + \frac{1}{2}\|x-u\|_P^2\), where we recall that, for all \(x\), \(\|x\|_P^2 = \langle x, Px \rangle\). In contrast to~\eqref{eq:PGD} where the preconditioning applies straightforwardly to the gradient, computing \(\operatorname{prox}_{\gamma f}^P(x)\) is more involved. In our case, we rely on a subiteration solver.

Following~\cite{hong_provable_2024}, we investigate three choices for matrix \(P\).
Denoting \(A \stackrel{\text{def}}{=} \mathcal{F}_\Omega \otimes \mathcal{S}_\ell\)
we consider the ``F-1'' preconditioner \(P = 2-\alpha A^\mathcal{H}A\), and the Chebychev preconditioner \(P = 4-\frac{10}{3} A^\mathcal{H}A\).
In a nutshell, the choice of these preconditioners arises from finding matrices \(P\) minimizing the spectral radius of \(\operatorname{Id}-\alpha P A^\mathcal{H}A\), see~\cite{hong_provable_2024} for more details. Lastly, we also investigate the case \(P=\operatorname{Id}\) corresponding to the standard PnP approach.
\begin{figure*}[htb]
  \centering
    \includegraphics[width=\textwidth]{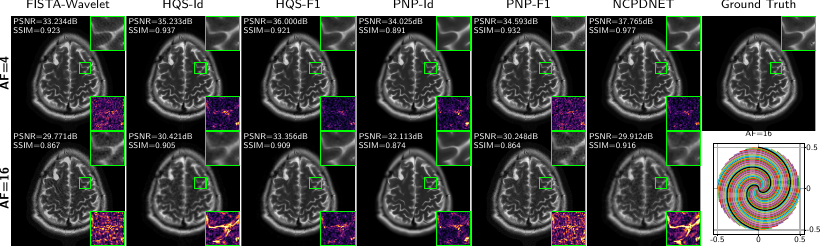}
   \caption{Top row: Reconstruction results for~\eqref{eq:MRI_multicoil} for HQS and PNP algorithms with no (Id) or static (F1) Preconditioner at AF=4 (top) and AF=16 (bottom). PSNR/SSIM metrics are shown inside each image. Bottom right inset depicts the residual maps (5\(\times\) magnified).}
  \label{fig:results}
\end{figure*}

\section{Experimental results and discussion}

\subsection{Simulated non-Cartesian acquisition}
Our ground-truth data consists of raw, fully-sampled multi-coil brain k-space data from the FastMRI dataset. To simulate an accelerated acquisition, we used a spiral pattern with either AF=4 or AF=8~(resp., 4- or 8-fold under-sampling). On the resulting under-sampled k-space, we added white Gaussian white noise (\(e_l\) in~\eqref{eq:MRI_multicoil}) with variance \(\nu_l=10^{-4}\cdot\max(\sum_{\ell}|x_\ell|^2)\).
The spatial sensitivity maps \((\mathcal{S}_\ell)_{1\leq \ell \leq L}\) in~\eqref{eq:MRI_multicoil} were estimated from the center of the k-space (20\(\times\)20 Hamming window), transformed in the image domain, and we masked-out the resulting images to fit the brain.

\subsection{Training of the denoising prior}

We trained our denoising prior \(\operatorname{D}_\sigma\) on the preprocessed data set \(\mathcal{D}_\text{train}\) at hand from Section~\ref{sect:dataset_gen}.
Following standard practice from the PnP literature, we chose a DRUNet model~\cite{zhang2021plug} conditioned on the noise level \(\sigma\) and with input and output convolutions of the model containing 2 channels accounting for real and imaginary parts. Additionally, the model was trained in a supervised way with an \(\ell_1\) loss on the proposed preprocessed dataset.
Training was performed for 100 epochs for noise levels \(\sigma\) ranging in \([0, 5\cdot 10^{-5}]\), with the Adam optimizer with learning rate \(10^{-4}\), with batch of size 10 and on randomly extracted patches of size 128\(\times\)128.
In contrast to other work in the MRI community where denoisers were trained on normalized images in \([0, 1]\), we proceeded differently as our denoisers worked on unnormalized MRI slices.
\begin{table}[!htb]
  \centering
  \vspace{2em}
  \caption{Reconstruction metrics for different solvers and acceleration factors. Reported values are an average of 5 trials. \textbf{First} and \underline{second} best performances are highlighted per AF (column-wise), across all PnP-like methods. Hyperparameters were chosen/ NCPDNET was trained at AF=4.\label{fig:metric_table}}
  \begin{tblr}{
width=\textwidth,
llrrrr,
cell{3}{1}={r=2,c=1}{c},
cell{5}{1}={r=2,c=1}{c},
cell{7}{1}={r=1,c=2}{l},
cell{8}{1}={r=1,c=2}{l},
cell{9}{1}={r=2,c=1}{c},
cell{11}{1}={r=2,c=1}{c},
cell{1}{3}={c=2}{c},
cell{1}{5}={c=2}{c},
cell{1}{1}={r=2}{c},
cell{1}{2}={r=2}{c},
hline{2}={3-4}{leftpos=-1,rightpos=1, endpos=true},
hline{2}={5-6}{leftpos=-1,rightpos=1, endpos=true},
colsep=3pt,
}
\toprule
\rotatebox{90}{Prec.} & Solver & PSNR  & & SSIM &   \\
  & Name & AF=4 & AF=16 & AF=4 & AF=16\\
\midrule
\rotatebox{90}{Cheb}
 & HQS & \textrm{34.521} & \underline{33.251} & \textrm{0.917} & \textrm{0.902}\\

 & PNP & \textrm{33.025} & \textrm{28.223} & \textrm{0.908} & \textrm{0.812}\\
\midrule
\rotatebox{90}{F1}
 & HQS & \textbf{35.874} & \textbf{33.641} & \textrm{0.921} & \textbf{0.911}\\

 & PNP & \textrm{34.596} & \textrm{30.280} & \underline{0.932} & \textrm{0.865}\\
\midrule
 PNP-FISTA & & \textrm{27.493} & \textrm{29.657} & \textrm{0.791} & \textrm{0.808}\\
\midrule
 PNP-Dyn & & \textrm{32.695} & \textrm{28.111} & \textrm{0.903} & \textrm{0.809}\\
\midrule
\rotatebox{90}{Id}
 & HQS & \underline{35.245} & \textrm{30.437} & \textbf{0.937} & \underline{0.905}\\

 & PNP & \textrm{34.043} & \textrm{32.118} & \textrm{0.892} & \textrm{0.874}\\
\midrule
\rotatebox{90}{N/A}
 & FISTA-Wavelet & \textrm{33.230} & \textrm{29.753} & \textrm{0.923} & \textrm{0.867}\\

 & NCPDNET & \textit{37.765} & \textit{29.912} & \textit{0.977} & \textit{0.916}\\
 \bottomrule
\end{tblr}

\end{table}

\subsection{Image reconstruction results}

\subsubsection{Preconditioning boosts image quality}
To evaluate the quality of our reconstructions, we use the Peak Signal-to-Noise Ratio (PSNR) and Structural Similarity Index (SSIM) as our primary metrics. Both metrics are computed on the magnitude of the output images, and the PSNR is restricted to the brain region of interest to avoid bias from the background.
We show in Fig.~\ref{fig:results} reconstruction results for AF=4 and AF=16 for PnP reconstructions. As in~\cite{hong_provable_2024} we observe that preconditioning significantly improves image quality compared to non-accelerated counterparts~(HQS-Id and PnP-Id). The visual improvements are in line with the broader quantitative assessment in Tab.~\ref{fig:metric_table}. Moreover, the HQS splitting scheme improves reconstruction over the standard PGD and FISTA PnP algorithms.
\subsubsection{Generalization to other acceleration factors}

The fully unrolled NCPDNET approach\cite{ramzi2022nc} still outperforms the PnP and HQS methods, when trained for a specific AF. Yet as the AF increases, the PnP methods shine, as they can maintain good image quality, without requiring retraining or fine-tuning a neural network. At AF=16 (Tab.~\ref{tab:metric_table}), all PnP methods outperform NCPDNet and variational approaches, without further fine-tuning (for each method, the same hyperparameters were used across all AF).

\begin{figure}[hbt]
    \centering
    \includegraphics[width=0.4\textwidth]{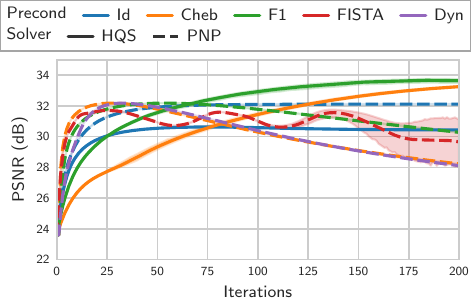}
    \caption{Evolution of the PSNR across iterations for various PnP algorithms at AF=16. Dashed lines correspond to variations of the~\eqref{eq:pnp_pgd} algorithm, while solid lines correspond to variations of the~\eqref{eq:dpir} algorithm.}
    \label{fig:convergence}
\end{figure}

\subsubsection{Stability of PnP methods and further acceleration}

We show in Fig.~\ref{fig:convergence} the convergence plots for all PnP and HQS algorithms at AF=4 Interestingly, we notice that the HQS algorithm is slower than its PGD counterpart with preconditioning, which indicates that further fine-tuning of the hyperparameters could lead to faster results. In particular, the stepsizes could be adaptively learned on a validation set.
Concerning the preconditioned PnP (PnP-F1, PnP-Cheb, PnP-Dyn)\cite{hong_provable_2024}, we did use a normalization equivariant network. Surprisingly, we were unable to reproduce the convergent behavior for those methods as shown in Fig.~\ref{fig:convergence}. In contrast, we relied on the annealing scheme for the HQS iterations to improve image quality.
 
\section{Conclusion \& Future Work}

In this work, we introduced a novel algorithm for multicoil, non-Cartesian MRI reconstruction by leveraging an annealed preconditioned Half-Quadratic Splitting (HQS) algorithm within the Plug-and-Play (PnP) framework. To avoid the limitations posed by the absence of high-quality, noise-free training data, we proposed an unsupervised preprocessing pipeline that effectively denoises multicoil k-space data. This preprocessing involves a virtual coil combination followed by training an unsupervised neural network for complex MRI signal denoising, generating a clean dataset suitable for subsequent training of a denoiser.

By training a high-performance denoiser on these clean data and integrating them into various PnP algorithms, we demonstrated the superior reconstruction quality of our approach compared to existing methods. Further work includes extension to 3D and 2D+time data, such as those considered in cardiac imaging or functional MRI.

Detailed settings, implementations, and a reproducible benchmark are also made available along this paper:

\footnotesize{\url{https://github.com/paquiteau/benchmark_mri_pnp}}

\section{Compliance with Ethical Standards}
This is a numerical simulation study for which no ethical approval was required.
\section{Acknowledgments}
This work was granted access to IDRIS' HPC resources under the allocation 2023 AD011011153R3 made by GENCI. The authors have no relevant financial or non-financial interests to disclose.
\printbibliography
\end{document}